\documentstyle[12pt,aasms4,epsf]{article}
\begin{document}

\title{Does the AGN Continuum Shape Change with Luminosity?}
\author{P.\ Romano and  B.M.\ Peterson}
\affil{Department of Astronomy, Ohio State University, \\
	174 West 18th Avenue, Columbus, OH 43210-1106, USA \\
	Email: promano, peterson@astronomy.ohio-state.edu}

       \begin{abstract}

We test the general wisdom that AGN continua  
vary with larger amplitude at shorter wavelengths, 
through a re-analysis of historical data from the optical 
and UV ({\it IUE\,}) 
 monitoring campaign on the Seyfert 1 galaxy NCG~5548 by 
the AGN Watch Consortium. 
We assume that the only non-varying component of the optical 
continuum is the integrated light from the host 
galaxy, which does not affect the UV continuum. 
Therefore, we expect any curvature in a linear plot of the 
UV continuum flux 
versus the simultaneous optical flux to represent a true change 
in the spectral index $\alpha$; the $y$-intercept   
provides an estimate of the host galaxy contribution in the 
optical region. 
We measured the continuum fluxes at 1350\,\AA{} (rest frame) 
from  the NEWSIPS-extracted spectra and adopted the optical 
continuum determinations of Wanders \& Peterson (1996). 

The data confirm that the curvature in the $F_{\lambda}$(5100) -- 
$F_{\lambda}(1350)$ plot is real, and that the spectral index is a 
function of luminosity. 
Hence for NGC~5548, the continuum does get harder as it gets brighter.
However, the relationship between the optical and UV continuum flux 
is more complicated than the models we tested would indicate.

	\end{abstract}

	\section{Introduction}

Among the early findings about AGN continua was the trend for larger 
amplitude variations to be observed at shorter wavelengths, 
so that the continuum becomes harder as it becomes brighter. 
This has been repeatedly claimed for NGC~5548 (\cite{Wamsteker90}, \cite{Clavelea91}, 
\cite{Peterson91}, \cite{Koristaea95}), NGC~3783 (\cite{Reichertea94}) 
and NGC~4151 (\cite{Crenshawea96}).
However,  Fairall~9,  at the high-luminosity end of the monitored 
sources, has been shown to have a wavelength-independent amplitude of continuum 
variations (\cite{Rodriguezea97}; \cite{Santosea97}). NGC~4593 has shown the same 
behaviour (\cite{Santosea95}).
Hence, we decided to reconsider the issue, especially in light of the ready 
availability of the NEWSIPS-extracted data from {\it IUE}\, for a large sample 
of well-monitored objects   
(now comprising NCG~5548, NCG~3783, NCG~7469, NGC~4151, Fairall~9, 3C390.3, 
and Mkn~509).   
We will adopt as a null hypothesis that the UV/optical continuum  in NGC~5548 
retains its shape as the luminosity varies.

	\section{Data Samples and Analysis}

We formed quasi-simultaneous (to within 3 days) pairs of UV/optical 
continuum measurements from observations made  between 1988 and 1993, 
the first five years of the International 
AGN Watch (\cite{Alloinea94}) spectroscopic monitoring campaign on NGC~5548.
For the UV continuum determinations we retrieved the {\em International Ultraviolet 
Explorer} Satellite ({\em IUE}\,) short wavelength prime (SWP) camera spectra 
taken during the optical campaign, as processed with 
the new {\em IUE}\, Final Archive software NEWSIPS (\cite{Nicholsea93}), 
which includes corrections for non-linearity that might have affected 
previous studies. 
The original data appeared in  Clavel et al.\ (1991), Clavel et al.\ (1992), 
and Korista et al.\ (1995). We determined the flux at 1350\,\AA{}
(rest frame) as the average flux over a 20\,\AA{} wide wavelength range 
around 1373\,\AA, and adopted  the uncertainties of $2\%$ of the 
flux values, as suggested by the statistical considerations in 
Rodr\'\i guez-Pascual et al.\ (1997), i.e., closely spaced determinations 
are independent measures of the same quantity, and therefore the pixel-based 
uncertainties from the  NEWSIPS software are overestimations.
 
We adopted the optical continuum determinations by Wanders \&\ Peterson (1996), 
where the optical continuum is the average over a 10\,\AA{} wide wavelength 
range centered 
around 5188\,\AA{} (5100\,\AA{} in the rest frame). The advantage over the previous 
determinations is that these constitute a more homogeneous subset of the original data 
(Peterson et al.\ 1991, 1992; \cite{Koristaea95}),  re-calibrated using the 
automatic scaling algorithm developed by \cite{vGW92}. 

\begin{figure}[b]
	\null
	\epsfxsize=13.5truecm
	\epsfysize=13.5truecm
	\vspace{-6.5truecm}
	\hspace{-0.5truecm}
	\epsfbox{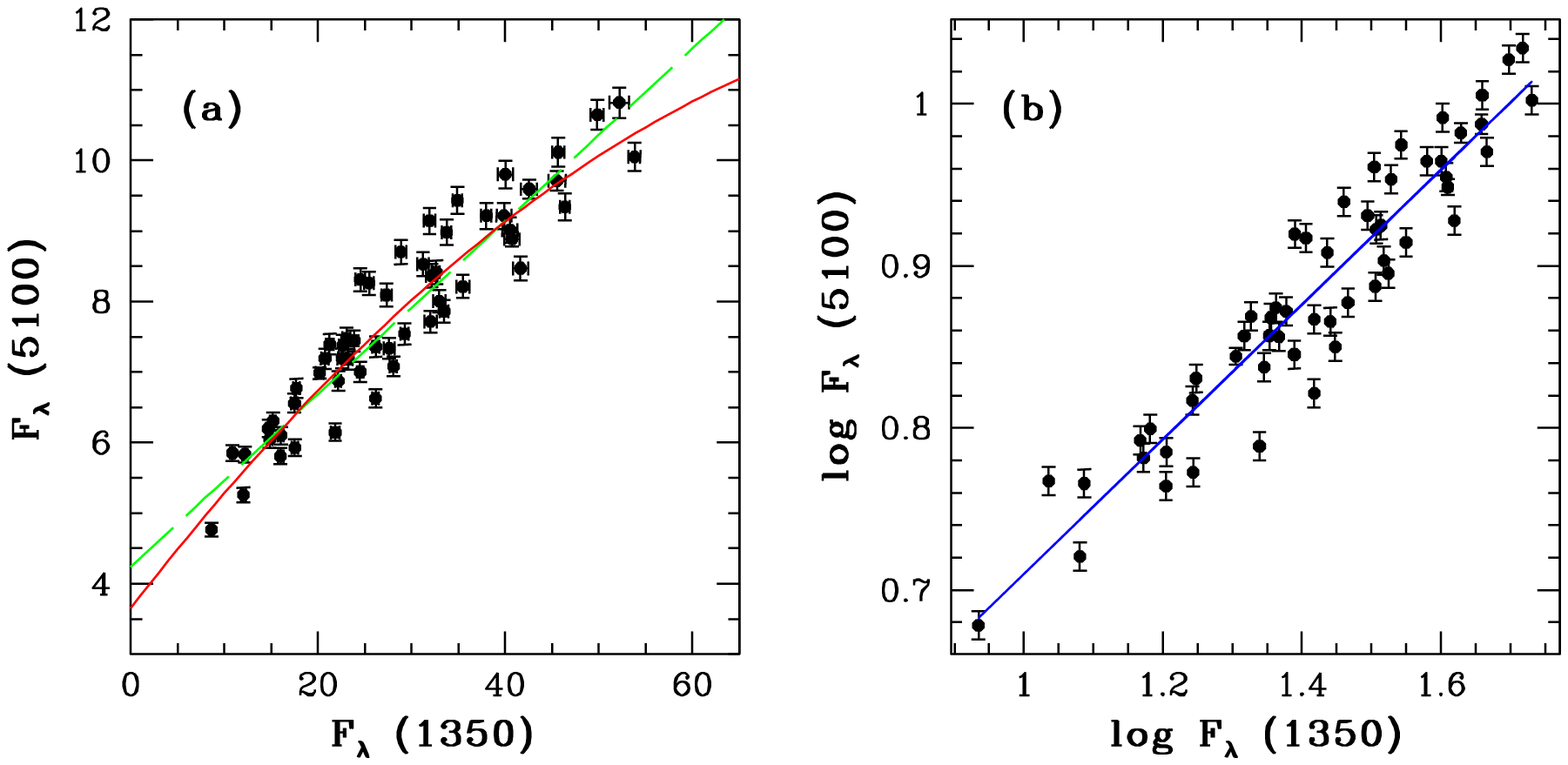}
	\caption[New Data]{\label{fig:newuvopt} Optical from \cite{WP96} vs.\ 
	{\it IUE}\, (NEWSIPS) data.   
	Fluxes are in units of 10$^{-15}$\,erg\,s$^{-1}$\,cm$^{-2}$\,\AA$^{-1}$. 
	Left: the polynomial fit (solid line); 
	linear fit (dashed line). 
	Right: the power law fit.  
	} 
\end{figure}

We are implicitly assuming that the only non-varying component of the optical 
continuum is the integrated light from the host galaxy, which does not 
affect the UV flux (cf \cite{Peterson91}). Hence, there is a simple way to test 
whether the continuum shape changes with luminosity. 
In a linear plot of the UV continuum flux versus the simultaneous optical flux, 
any curvature represents a true change 
in the spectral index $\alpha$, and the $y$-intercept provides an estimate of the host galaxy contribution in the optical region. 
The first test we performed was for a simple functional form for the dependence of
the optical flux on the UV flux, namely, polynomial and power law. 
Figure~\ref{fig:newuvopt}a shows the best fit, the  second-order polynomial 
F$_{\lambda}(5100) =  3.65 \pm 0.35 +(0.17 \pm 0.03)\,F_{\lambda}(1350) 
-(8.60 \pm 4.67)\, \times 10^{-4} \,F_{\lambda}^2 (1350)$, reduced $\chi^2$,
$\chi^2_\nu = 9.91$; and the linear fit, 
$F_{\lambda}(5100) = 4.24 \pm 0.16 +(0.12 \pm 5.73 
\times 10^{-3})\,F_{\lambda}(1350)$, $\chi^2_\nu = 10.36$. 
Fluxes are in units of 10$^{-15}$\,erg\,s$^{-1}$\,cm$^{-2}$\,\AA$^{-1}$. 
The $y$-intercept in the second-order polynomial fit is $3.65 \pm 0.35$, which is 
consistent with the value $3.4 \pm 0.4$ by \cite{Romanishinea95} 
obtained from directing imaging of the source. 
However, the best fit is the  power law (Figure~\ref{fig:newuvopt}b), 
log $F_{\lambda} (5100) = 0.29 \pm 0.03 +(0.42 \pm 0.02)$ 
log $F_{\lambda} (1350)$, $\chi^2_\nu = 9.65$. 

We then considered the possibility that $F_{\rm gal}$ had been underestimented by \cite{Romanishinea95} thus making a true linear relationship a power law: 
$F_{\rm opt} = a + F_{\rm uv} ^{\gamma}.$ 
We sought for a value for the galaxy contribution to the optical that would 
give a value of $\gamma$ consistent with 1. Such a value is 
$F_{\rm gal} = (4.55 ^{+ 0.22}_{-0.15})$, and in that case the fit is: 
log $F_{\lambda} (5100) = -0.939 \pm 0.085 + (1.001 \pm 0.056)$ 
log $F_{\lambda} (1350)$, $\chi^2_{\nu} = 9.36$. 
The upper limit on $F_{\rm gal}$ is given by the condition that the maximum of the galaxy 
contribution cannot exceed the minimum observed flux, which occurred on JD 2,448,810.

\begin{figure}[p]
	\null
	\epsfxsize=13.5truecm
	\epsfysize=16truecm
	\vspace{-0.5truecm}
	\epsfbox{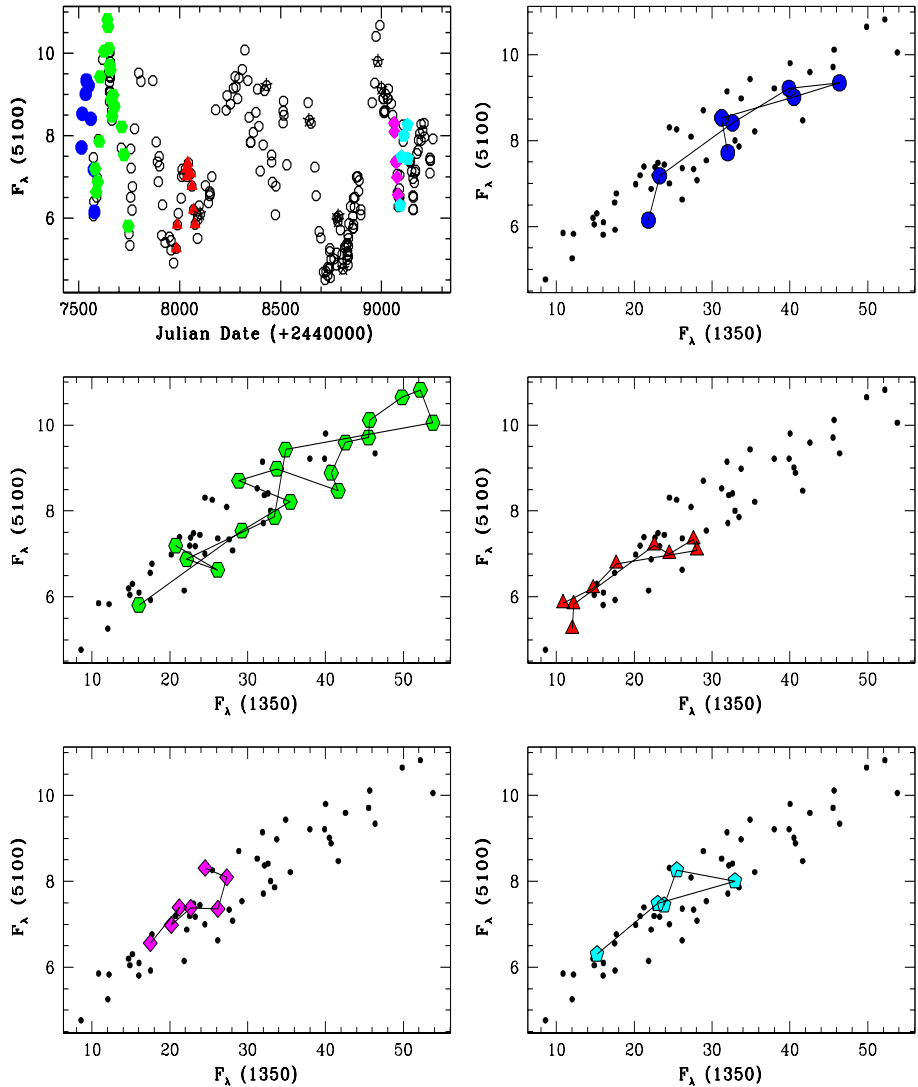}
	\caption[JDordered]{\label{fig:jdordered} 
	Panel {\bf (a)}: optical light curve for NGC 5548. 
	Panels {\bf (b)} to {\bf (f)}: 
	{\it IUE}\, (NEWSIPS) data vs.\ optical data connected according 
	to Julian Date during each activity event. 
	Fluxes are in units of 10$^{-15}$\,erg\,s$^{-1}$\,cm$^{-2}$\,\AA$^{-1}$.
	} 
\end{figure}

As a last test for linearity, we considered the possibility that the 
spread observed in the points in Figure~\ref{fig:newuvopt}a 
and the resulting curvature, might be due to the superposition of subsets 
of the whole sample, which might be otherwise fit linearly. 
The natural choice for these subsets is represented by activity events. 
In the top panel of Figure~\ref{fig:jdordered}, the empty circles represent 
the UV light curve, and the filled ones have matching optical points. 
They have been divided into 5 subsets, according to the Julian Date, 
and fit individually. In none of the cases is the 
linear fit the best. Indeed, events 1, 3, and 5 are best fit by 
a second-order polynomial and events 2 and 4 by power laws.

\begin{figure}[p]
	\null
	\epsfxsize=13.5truecm
	\epsfysize=12truecm
	\vspace{-0.5truecm}
	\epsfbox{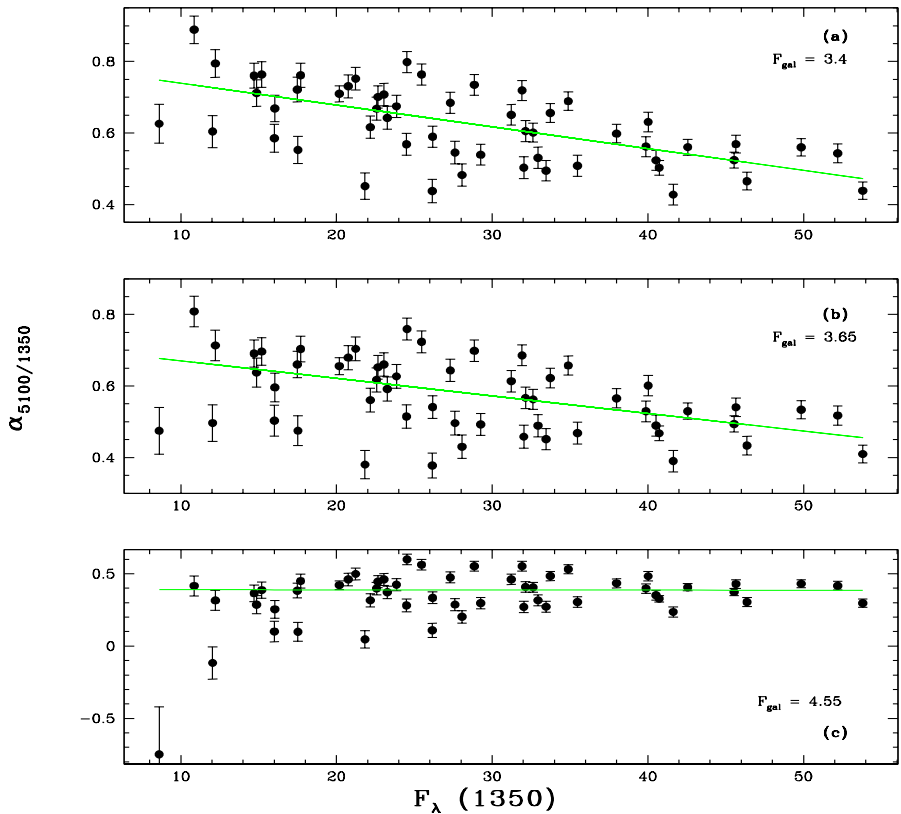}
	\caption[Alpha Fits UV OPT]{\label{fig:alphaUVOPT} 
	Correlation between the spectral index $\alpha_{5100/1350}$ and the UV 
	flux $F_{\lambda} (1350)$.
	Panel {\bf (a)}: For a galaxy contribution to the optical flux of 
	$F_{\rm gal} = (3.4 \pm 0.4) \times 
	10^{-15}$\,erg\,s$^{-1}$\,cm$^{-2}$\,\AA$^{-1}$;
	Panel {\bf (b)}: For $F_{\rm gal} = (3.65 \pm 0.35) \times 
	10^{-15}$\,erg\,s$^{-1}$\,cm$^{-2}$\,\AA$^{-1}$;
	Panel {\bf (c)}: For $F_{\rm gal} = (4.55 ^{+ 0.22}_{-0.15}) \times 
	10^{-15}$\,erg\,s$^{-1}$\,cm$^{-2}$\,\AA$^{-1}$. 
	}
\end{figure}

Assuming the power law shape 
$F_{\lambda} = k \, \lambda^{\alpha -2}$  
for the continuum,  we calculated the spectral index $\alpha$
for three values of the galaxy contribution to the optical flux, namely 
(in units of 10$^{-15}$\,erg\,s$^{-1}$\,cm$^{-2}$\,\AA$^{-1}$), 
(a) $3.4 \pm 0.4$ from \cite{Romanishinea95}; 
(b) $3.65 \pm 0.35$ from our second-order polynomial fit; 
and (c) $4.55 ^{+ 0.22}_{-0.15}$, from the power law fit for $\gamma =1$.  
There is a statistically significant slope in all cases 
(Figure~\ref{fig:alphaUVOPT}): 
	(a) $\alpha_{5100/1350} = 0.800 \pm 0.033 -(6.09 \pm 0.985) 
\times 10^{-3}$  F$_{\lambda}(1350)$; 
	b) $\alpha_{5100/1350} = 0.720 \pm 0.035 -(4.92 \pm 0.105) 
\times 10^{-3}$ F$_{\lambda}(1350)$;
	(c) $\alpha_{5100/1350} =  0.391 \pm  0.044 -(9.94 \pm 138.) 
\times 10^{-5}$ $F_{\lambda}(1350$).
This proves that $\alpha$ is not constant as the flux varies, that is, 
$\alpha$ is a function of luminosity, so that for NGC~5548, 
the continuum does get harder as it gets brighter.

	\section{Conclusions and Summary}

We have compared contemporaneous  optical and UV data from the first 5 years 
of the spectroscopic monitoring campaign on NGC~5548, and conclude the 
amplitude of variation is greater at shorter wavelength, i.e.,  
the continuum does get harder as it gets brighter. 
We have shown that the curvature in the optical/UV plot is real and that 
the  spectral index $\alpha$ depends on luminosity. 
However, the high values of $\chi^2_{\nu}$ for each fit imply that the models 
tested are not adequate in describing our data. Therefore, the relationship 
between the optical and the UV continuum flux is probably better described by 
a functional form that is more complicated than either polynomial or power law.
The logical next step of the work is the extension to the other well-monitored 
sources, NCG~3783, NCG~7469, NGC~4151, 3C390.3, Mkn~509, and Fairall~9.
In particular this last object that has a wavelength-independent amplitude of 
continuum variations.

\end{document}